# Selective epitaxial growth of graphene on SiC

N. Camara,[1] G. Rius,[1] J.-R. Huntzinger,[2] A. Tiberj,[2] N. Mestres,[3] P. Godignon[1] and J. Camassel[2]

[1]*CNM-IMB-CSIC – Campus UAB 08193 Bellaterra, Barcelona, Spain*
[2]*GES - UMR 5650 Université Montpellier 2/CNRS, 34095 Montpellier cedex 5, France*
[3]*ICMAB-CSIC, Campus UAB 08193 Bellaterra, Barcelona, Spain*

We present an innovative method of selective epitaxial growth of few layers graphene (FLG) on a "pre-patterned" SiC substrate. The methods involves, successively, the sputtering of a thin AlN layer on top of a mono-crystalline SiC substrate and, then, patterning it with e-beam lithography (EBL) and wet etching. The sublimation of few atomic layers of Si from the SiC substrate occurs only through the selectively etched AlN layer. The presence of the Raman G-band at ~1582 cm$^{-1}$ in the AlN-free areas is used to validate the concept, it gives absolute evidence of the selective FLG growth.

Graphene is a 2-dimensional carbon system with outstanding, pseudo-relativistic, transport properties[1]. Once properly processed, the graphene-based devices will be candidates to breakthrough the Si leadership in C-MOS microelectronic industry for RF applications. A widely used technique to fabricate mono or bi-layers graphene flakes is the exfoliation of HOPG (Highly Oriented Pyrolytic Graphite) on an oxidized Si wafer. The CVD growth of carbon on metals like Ru(0001)[2] have also been considered. However, up to now, the best way to fabricate FLG (Few Layer Graphene) on a full wafer for industrial purpose is high temperature sublimation of few atomic layers of Si from a mono crystalline SiC substrate[3]. Then, the first step in graphene-base device technology is to form FLG nanoribbons by performing dry etching in an $O_2$ plasma. This step introduces defects and dangling bonds at the edges of graphene nanoribbons, decreasing their carriers mobility[4]. Also, the PMMA and the chemical wet solutions used for patterning induce an unintentional doping in the graphene sheets, altering the device performance[5].

The main purpose of this work is to show that micro-size FLG can be produced through a sputtered patterned AlN layer on top of a SiC sample. In addition of an obvious relevance for industrial purpose, one of the main advantages of this technique is that the future dielectric layer used as top gate for the FET-like device can be deposited immediately, right after the graphene growth over the full wafer. In such a way, the graphene ribbon never "sees" any chemical for the rest of the device process. This advantage is relevant only for the non sensing devices.

AlN is currently used in SiC device technology as capping layer, protecting the implanted SiC surface during post-implantation annealing at high temperature[6,7]. It is very stable at temperatures SiN or $SiO_2$ can not sustain and no reaction occurs between the AlN layer and the SiC substrate up to 1650°C[6]. It is also very easy to etch in a wet TMAH (TetraMethylAmmonium Hydroxide) solution, which is also extremely selective and does not react with the SiC surface. This is very important since the surface has to remain atomically flat before Si-sublimation. Finally, TMAH is C-MOS clean room compatible, a crucial point for the industrial application of this technology. Altogether, this makes AlN the best candidate for any "pre-patterned" SiC technology.

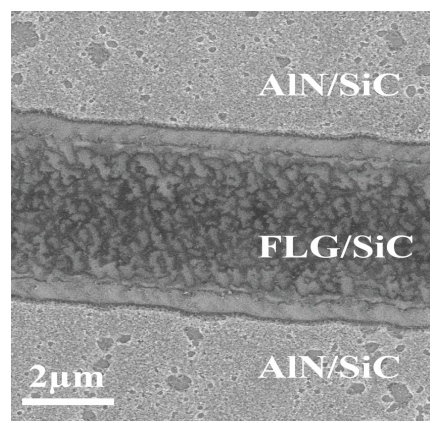

FIG. 1. SEM image of FLG ribbon grown on SiC using a patterned AlN mask.

In this work, two pieces of C-face 6H-SiC, on-axis, were cut from a n-type wafer, doped to ~5x10$^{17}$ cm$^{-3}$. The wafer was epi-ready polished by Novasic up to an atomically flat surface. The C-face orientation was chosen because previous experiments showed that FLG grown on this orientation is more promising in terms of domain size and smoothness[8]. To guarantee C-MOS compatibility, all chemical treatments before Si sublimation were RCA and Piranha cleaning. Then, a thin 100nm AlN layer was sputtered on one of the 2 samples, patterned by EBL and etched in TMAH down to the SiC substrate. The patterned ribbons were 4μm wide and 100 μm long. Si-sublimation was done simultaneously on the two samples, in a commercial RF furnace from JIPELEC. The vacuum was

~ $10^{-6}$ Torr. After few preliminary steps at 1050°C for 10 min in order to remove any trace of native oxide, the samples were heated at 1550°C (as controlled by a pyrometer) during 5 min.

On sample A with no AlN mask on top, after Si-sublimation at 1550°C, the whole surface was covered with FLGs, presenting brighter flakes visible by optical microscopy and SEM (Scanning Electron Microscopy). The surface topography measured by AFM (Atomic Force Microscope) revealed a step bunched surface with atomically flat domains of ~ 500 nm length for the largest ones. This is in the range of literature data[9].

On sample B with patterned AlN on top, SEM measurements revealed that the AlN mask did not suffer from the 1550°C sublimation step (Fig. 1). However, the graphitization result looked different. The FLG domain bounderies were no longer visible by AFM and only Raman spectroscopy could identify the graphitized part inside the narrow ribbons.

In order to compare the FLG grown on sample A (un-patterned) and sample B (AlN patterned), we used Raman spectroscopy. Micro-Raman spectra were collected through a x100 objective at room temperature, in the back-scattering configuration [10-12] using the 514.5 nm (2.41 eV) line of an Ar+ laser. The incident power on the sample was 1 mW. Internal frequency and normalization calibration were provided by the second order modes of the SiC substrates. The highest peak of the SiC second order modes at 1516 $cm^{-1}$ was taken as Raman shift reference, its intensity being used to estimate the number of FLGs and normalize the spectra.

For both samples (A and B) we resolved three bands: the G-band at ~ 1582 $cm^{-1}$, then the 2D-band at ~ 2700 $cm^{-1}$ and, finally, the D-band at ~ 1350 $cm^{-1}$. The G-band originates from the degenerate phonon mode ($E_{2g}$) at the centre of the Brillouin Zone. This mode is characteristic of graphitic compounds whatever is their stacking order or their crystallite size. In Fig. 2 we find clear evidence of this band on the Raman spectrum of samples A and B (with and without AlN masking). No G-band is found on sample B below the AlN mask. This proves unambiguously that the concept of patterned FLG epitaxy through an AlN mask is sustainable.

Similar to previous work[10-12], we find that the G peak position is up-shifted with respect to exfoliated graphene on oxidized Si. This substrate effect is not yet perfectly understood but recent results show that the higher the number of layers, the lower the frequency of the G-band[12]. Combining intensity and frequency of the G-band, we find that the FLG grown on the AlN patterned chip is thinner than the one grown on the bare SiC sample. Thanks to previous Raman spectroscopy data[13], we estimate the number of layers from 10 to 15 layers for sample A and 5 to 10 layers for sample B. Such thicknesses are standard for FLG on SiC-based devices[9].

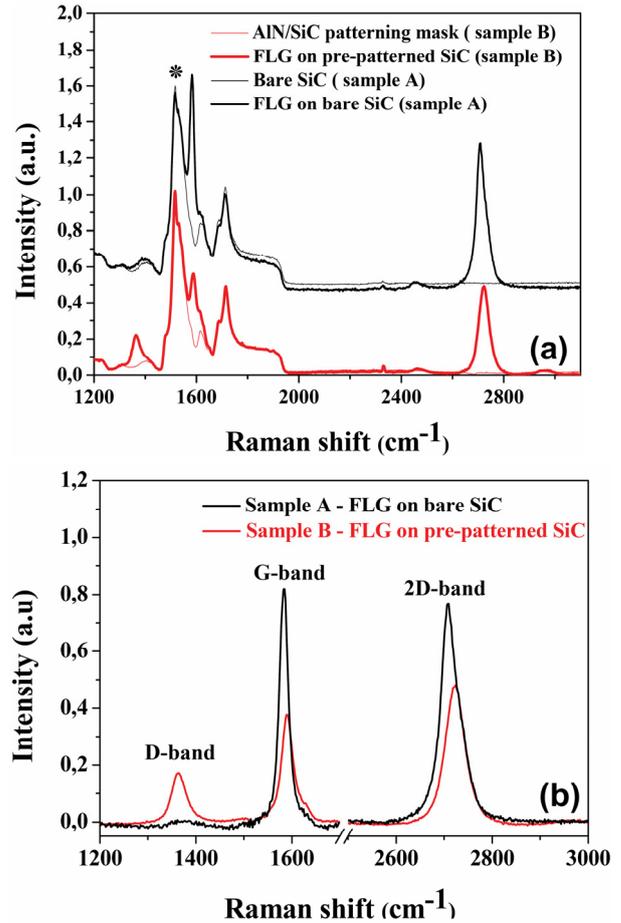

FIG. 2. Raman spectra of (a) FLG grown on bare SiC and FLG grown on "AlN patterned" SiC. The SiC reference spectra are superimposed, the reference spectra was done before the Si-Sublimation on sampleA while the reference was done outside the ribbon, after Si-sublimation on sample B. (b) Raman spectra after SiC reference subtraction for both samples. * is the main second order SiC peak used to calibrate in frequency and to normalize the spectra.

The 2D band at ~ 2700 cm-1 comes from a two-phonons double resonance effect. It is very sensitive to the electronic structure[14] and changes with the stacking order around the c-axis[15]. In this way, it was found possible to follow the evolution from turbostatic (with only one wide peak) to HOPG (with 2 broad peaks) while heating up a graphitic systems[15]. From our FLGs, with or without AlN masking, we observe the typical signature of the lack of long-range order between adjacent layers. This indicates turbostratic like graphite with a single broad 2D band and FHWM of ~ 42 $cm^{-1}$ [10-12,15](see Table I). For sample A, the 2D band Raman shift is 2711 $cm^{-1}$ while for sample B the 2D band Raman shift is 2722 $cm^{-1}$ (see Fig. 2 and Table I). Again, the difference in frequency and intensity comes from the FLG thickness difference. Compressive strain has been invoked to explain this shift. This strain might be due to lattice mismatch[11] or differential thermal dilation when the sample cools down[12], between the FLG and SiC. In both cases, it is assumed that the thicker the FLG, the lower the 2D band frequency. Indeed, the FLG are more relaxed since the influence of the substrate decreases. Several groups have already shown evidences of disordered stacks on epitaxial FLG grown on SiC either electrically[9], by X-ray diffraction[16] or by micro Raman microscopy[10]. It was also shown that, thanks to this lack of order between adjacent layers, FLG on SiC exhibits the well known linear dispersion at the K point of the Brillouin zone, even when the domain sizes are as low as 120nm length[9,17].

TABLE I. D, G and 2D Raman shift and FWHM of samples A and B as well as the average domains length.

| Sample | D-band (cm$^{-1}$) | | G-band (cm$^{-1}$) | | 2D-band (cm$^{-1}$) | | L$a$ (nm) |
|---|---|---|---|---|---|---|---|
| | Δω | FWHM | Δω | FWHM | Δω | FWHM | |
| A | 1364 | 35 | 1587 | 28 | 2711 | 44 | 160 |
| B | 1376 | 84 | 1590 | 27 | 2722 | 45 | 30 |

The domain size can be estimated from the intensity of the D band at ~ 1350 cm$^{-1}$. It is due to a defect-assisted one-phonon double resonance effect and reflects the size of the FLG domain or the crystalline defects[18]. Comparing the Raman integrated intensity ratio of the D and G bands in samples A and B, we evaluate the domain size of our FLGs. We followed the empirical path of Pimenta et al[15]:

$$L_a = \frac{560}{E_l^4}\left(\frac{I_G}{I_D}\right) \quad (1)$$

in which $L_a$ (nm) is the average domain size, $E_l$ (eV) the energy of the laser beam and $I_D$, $I_G$ the integrated intensities of the D and G bands, respectively. The ration $I_D/I_G$ of the FLG without AlN mask is about 1/10, corresponding to an average crystallite size $L_a$ of ~160 nm. This is close to the most recent FLGs presented in literature (~210 nm)[9]. The domain size estimated this way is a lower limit on the actual size of the domain size when comparing to our AFM measurements. The FLG grown through the AlN mask on sample B exhibits smaller domain sizes. The ratio $I_D/I_G$ for sample B being 3/5, we evaluate the FLG flakes size to be ~30 nm in average. This smaller domains size may be an intrinsic problem of having Al and N atoms in the furnace during the Si-sublimation process. It may also originate from a small degradation of the SiC surface during AlN sputtering. It may also come from a lack of appropriate treatment after AlN wet etching. Obviously, our technology would have to be optimized when looking at the fabrication of very high mobility FET transistors. However, in some cases, smaller domain size is not a handicap. This is already true for applications like electrochemical sensors in which polycrystalline graphite is currently used to measure redox reactions in the ranges that metallic (usually Pt or Au) electrodes cannot reach[19]. Our technique of AlN patterning in this case is of particular interest and directly suitable for interdigitated array microelectrodes for redox signal amplification[20].

To conclude, we have demonstrated that FLG can be selectively grown through openings in a sputtered AlN layer. No doubt exists about the nature of the FLG since, thanks to Raman signature, evidence of graphitic sp$^2$ bonding is found, with flakes dimensions in the range of 30nm in average and turbostratic arrangement. Our technique facilitates the graphene-based device process technology over the whole SiC wafer and opens the road to a new generation of mixed graphene-based / SiC-based devices on a common substrate.

This work was supported by the French ANR project GraphSiC, C-Nano Project and the Spanish grant Juan de la Cierva. The authors would like to thanks J. L. Sauvajol for the Raman spectroscopy equipement.